\documentclass[pre,aps,twocolumn,english,superscriptaddress,floatfix]{revtex4-1}
\usepackage[T1]{fontenc}
\usepackage[latin9]{inputenc}
\usepackage{xcolor}
\usepackage{amsmath}
\usepackage{amsthm}
\usepackage{amssymb}
\usepackage{graphicx}
\usepackage{subfigure}
\usepackage{braket}
\DeclareMathOperator{\Tr}{Tr}

\usepackage[strict]{changepage}

\makeatother

\usepackage{babel}

\begin{document}
\title{Thermodynamic Uncertainty Relations for Bosonic Otto Engines}

\author{Massimiliano F. Sacchi}
\email{msacchi@unipv.it}
\affiliation{CNR - Istituto di
  Fotonica e Nanotecnologie, Piazza Leonardo da Vinci 32, I-20133,
  Milano, Italy,}  \affiliation{QUIT Group, Dipartimento di Fisica,
  Universit\`a di Pavia, via A. Bassi 6, I-27100 Pavia, Italy.}

\begin{abstract}
  We study two-mode bosonic engines undergoing an Otto cycle.  The
  energy exchange between the two bosonic systems is provided by a
  tunable unitary bilinear interaction in the mode operators modeling
  frequency conversion, whereas the cyclic operation is guaranteed by
  relaxation to two baths at different temperature after each
  interacting stage.  By means of a two-point-measurement approach we
  provide the joint probability of the stochastic work and heat. We
  derive exact expressions for work and heat fluctuations, identities
  showing the interdependence among average extracted work,
  fluctuations and efficiency, along with thermodynamic uncertainty
  relations between the signal-to-noise ratio of observed work and
  heat and the entropy production. We outline how the presented
  approach can be suitably applied to derive thermodynamic uncertainty
  relations for quantum Otto engines with alternative unitary strokes.
\end{abstract}
\maketitle
\section{Introduction}
Nonequilibrium processes are always accompanied by irreversible
entropy production \cite{prig}. When systems become smaller, as in
nanoscopic heat engines \cite{nano1,nano2}, biological or chemical
systems \cite{gnes,rit,rao} or nanoelectronic devices
\cite{ventra,soth}, the fluctuations of all thermodynamic quantities
as work, heat, their correlations, and entropy production itself,
become very relevant. For example, a macroscopic thermal engine
supplies a certain amount of work while extracting heat from a hot
thermal reservoir. As the thermodynamic machine size is reduced, the
work output and heat absorbed are correspondingly scaled down, their
fluctuations become more and more significant, and it becomes useful
to investigate the stochastic properties of such fluctuating
quantities.  \par A number of fluctuation theorems has been derived
\cite{evans,gal,jar97,crook,piecho,jarz,jarz3,seif2,marc,saito,andrie,esp,esp2,cth,sini,jarz2,camp,seif,frq,hang,esp3}
as powerful relations that characterize the behavior of small systems
out of equilibrium. Fluctuation relations pose stringent constraints
on the statistics of fluctuating quantities as heat and work due to
the symmetries (particularly, time-reversal symmetry) of the
underlying microscopic dynamics.  Furthermore, recent relations have
also been developed, so called thermodynamic uncertainty relations
(TUR), where the signal-to-noise ratio of observed work and heat has
been related to the entropy production
\cite{bar,pietz,ging2,pole,pietz2,horo,proes2,agar,koy,bar2,brad,piet,holu,
  macie,Li,sary,dech,proes,bar3,guar,ging}.  Such TURs rule for
example the tradeoff between entropy production and the output power
relative fluctuations, i.e. the precision of a heat machine, so that
working machines operating at near-to-zero entropy production cannot
be achieved without a divergence in the relative output power
fluctuations.

\par Although independently developed, fluctuation relations and TURs
have been recently connected under various approaches and assumptions
\cite{proes2,merh,vanvu,potts,vanvu2,timpa,zhang,vanvu3}.  In
particular, in Ref. \cite{timpa} a saturable TUR obtained from
fluctuation theorems has been derived and compared with exact results
pertaining to a microscopic two-qubit swap engine operating at the
Otto efficiency.

\par In this paper we derive thermodynamic uncertainty relations for
two-mode bosonic engines, where alternately each quantum harmonic
oscillator is coupled to a thermal bath allowing heat exchange, and a
unitary bilinear interaction determines energy exchange between the
two modes by frequency conversion with tunable strength. We adopt the
two-point-measurement scheme \cite{esp,der,th,camp} usually considered
in the derivation of Jarzynski equality \cite{j97} and referred to the
simultaneous estimation of both work and heat in order to derive the
joint characteristic function that provides all moments of work and
heat.  The model is shown to achieve the Otto efficiency
\cite{ot1,ot2,ot3,ot4,ot5,cpf,dec,ot6}, independently of the coupling
parameter and the temperature of the reservoirs.  After identifying the
regimes where the periodic protocol works as a heat engine, a
refrigerator, or a thermal accelerator, we provide the full
joint probability of the stochastic work and heat in closed form.
\par Our
derivation allows to obtain the exact relation between the
signal-to-noise ratio of work and heat and the average entropy
production of the engine, thus showing the deep interdependence among
average extracted work, fluctuations, and entropy production. From
these relations we derive thermodynamic uncertainty relations that are
satisfied in all the regimes of operations and for any value of the
bilinear coupling between the two quantum harmonic
oscillators. A bound of the efficiency in terms of the average work
and its fluctuations is also obtained.
\par As outlined in Appendix C, the presented approach can be applied to quantum 
thermodynamic engines with alternative 
unitary strokes in order to assess the validity of the standard TUR. 
\begin{figure}[htb]
\includegraphics[scale=0.91]{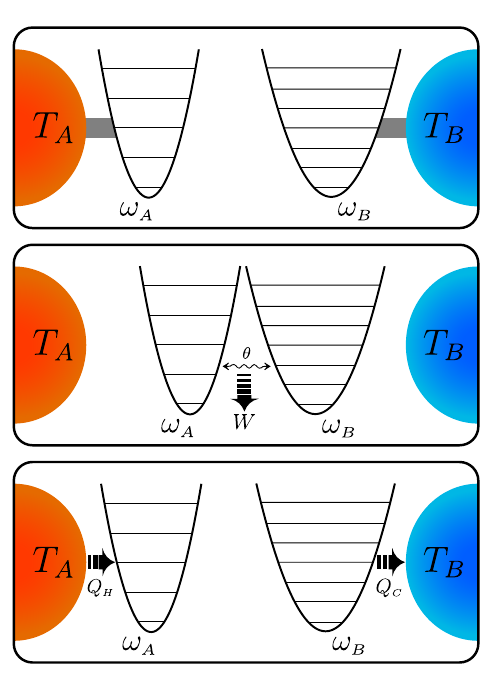}
\caption{Two-mode bosonic Otto cycle in heat engine
  operation: in the first stage each quantum harmonic oscillator with
  frequency $\omega _A$ and $\omega _B$ is in thermal equilibrium with
  its respective bath at temperature $T_A$ and $T_B$, respectively,
  with $T_A > T_B$; in the second stage the two oscillators are
  isolated and let to interact by a bilinear unitary interaction
  ($\theta $), thus extracting work $W$; in the third stage the
  oscillators are let to relax to their respective thermal baths, thus absorbing
  heat $Q_H$ and releasing heat $Q_C$, such
  that the initial condition is reestablished. In the refrigeration
  regime all three arrows are reversed.}
\end{figure}
\section{The two-mode bosonic Otto engine}
We illustrate now the two-mode bosonic engine under investigation, as
depicted in Fig. 1.  Let
us fix natural units $\hbar =k_B=1$. Each system is described by
bosonic mode operators $a,a^{\dagger}$ and $b,b^{\dag}$, respectively, with
the usual commutation relation, and corresponding free Hamiltonians
$H_{A}=\omega_{A}\left(a^{\dag}a+\frac{1}{2}\right)$ and
$H_{B}=\omega_{B}\left(b^{\dag}b+\frac{1}{2}\right)$.
Initially, the two modes $a$ and $b$ are in thermal equilibrium with
their own ideal bath at temperature $T_A$ and $T_B$, respectively, and we fix $T_A >
T_B$. Hence, the initial state is characterized by the tensor product
of bosonic Gibbs thermal states, i.e. 
\begin{eqnarray}
  \rho _0 = \frac{e ^{-\beta _A H_A}}{Z_A} \otimes \frac{e ^{-\beta _B H_B}}{Z_B} 
  \;,\label{eqq}
\end{eqnarray}
with $\beta _X = 1/T_X$ and $Z_X=\Tr [e^{-\beta _X H_X}]$. The two
systems are then isolated from their thermal baths and are allowed to
interact via a global unitary transformation.  We will consider the
bilinear interaction that globally transforms the mode operators as
follows
\begin{eqnarray}
&&a'=a \cos\theta +e^{i\varphi}b \sin\theta  ,\label{BS mode a-1}\\
&&b'=b \cos\theta - e^{-i\varphi}a \sin\theta  ,\label{BS mode b-1}
\end{eqnarray}
with $\theta \in [0,\frac{\pi}{2}]$ and $\varphi \in [0,2\pi]$.  \par
The Heisenberg transformations in Eqs. (\ref{BS mode a-1}) and
(\ref{BS mode b-1}) correspond to a linear mixing of the modes that
for $\omega _A \neq \omega _B$ describe frequency conversion, and in
the Schr\"odinger picture are equivalent to the unitary transformation
$U_ \xi =\exp\left ( \xi a^{\dag}b-\xi ^{*}ab^{\dag}\right )$, with
$\xi = \theta e^{i\varphi }$. We remark that $U_\xi $ incorporates the
free evolutions, all interactions and classical external drivings,
such that the corresponding unitary for the time-reversed process is
just $U^\dag _\xi$. We
also notice that an extensive study of such thermodynamic coupling,
especially for general Gaussian bipartite states, has been recently
put forward in Ref. \cite{bil}.  In a quantum-optical scenario, this
bilinear coupling may arise from an interaction Hamiltonian of
duration $t$ between the couple of modes $a$ and $b$ and 
a third mode at frequency $|\omega _A - \omega _B|$ considered as
a classical undepleted coherent pump with amplitude $\gamma $ via a
nonlinear $\chi ^{(2)}$ medium under parametric approximation
\cite{mand,par}, such that in the interaction picture $\xi = \gamma
\chi ^{(2)} t$.  In what follows the phase
$\varphi $ is irrelevant, hence we pose $\varphi =0$.  \par After the
interaction the two harmonic oscillators are reset to their
equilibrium state of Eq. (\ref{eqq}) via full thermalization by weak
coupling to their respective baths.  The procedure can be sequentially
repeated and leads to a stroke engine. We notice that for $\theta
=\pi/2 $ the unitary $U_{\pi/2}$ performs a swap gate which exchanges
the states of the two quantum systems, analogous to the two-qubit swap
engine \cite{cpf,timpa}. More generally, here we consider an arbitrary
value of $\theta $, modeling different interaction strengths (or
times). In each cycle the energy change in mode $a$ due to the unitary
stroke corresponds to the heat $Q_H$ released by the hot bath,
i.e. $Q_H = -\Delta E_a$, and similarly we have $Q_C= -\Delta E_b$ for
the heat dumped into the cold reservoir (heat is positive when it
flows out of a reservoir).  The work W is performed ($W>0$) or
extracted ($W<0$) during the unitary interaction, and from the first
law we have
\begin{eqnarray}
W=-Q_H -Q_C = \Delta E_a +\Delta E_b
  \;.
\end{eqnarray}
We can characterize the engine by the independent random variables $W$
and $Q_H$, and study the characteristic function $\chi(\lambda , \mu
)$, where $\lambda$ and $\mu$ denotes the work and heat labels such
that all moments of work and heat can be obtained by the identity
\begin{eqnarray}
  \langle W^n Q_H ^m \rangle = (-i)^{n+m}\left.
\frac{\partial  ^{n+m} \chi (\lambda ,\mu )}{\partial \lambda ^n
    \partial \mu ^m}\right
  |_{\lambda=\mu =0}
\label{mom}\;.
\end{eqnarray}
The characteristic function depends on the procedure that is adopted
to jointly estimate $W$ and $Q_H$. By using the two-point measurement scheme
\cite{esp,der,th,camp}, we can write the
characteristic function as follows \cite{camp}
\begin{eqnarray}
&&
\chi (\lambda ,\mu )= \label{6}
\\&&  \Tr [U^\dag _\theta e^{-i \mu H_A}e^{i \lambda (H_A +H_B)}
    U_\theta e^{i \mu H_A}e^{- i \lambda (H_A +H_B)} \rho_0 ]
    \;.\nonumber
\end{eqnarray}
By representing the thermal states as mixture of coherent states, namely
\begin{eqnarray}
  \frac {e^{-\beta _X H_X}} {Z_X}=
  \int \frac {d^2 \gamma } {\pi N_X} e^{-\frac {|\gamma
      |^2}{N_X}}|\gamma \rangle \langle \gamma |   
  \;,
\end{eqnarray}
with $d^2 \gamma = d \mbox{Re}\gamma \; d \mbox{Im}\gamma $ and
$N_X =
(e^{\beta _X \omega _X}-1)^{-1}$, 
from the identities
$e^{i\psi a^\dag a}|\alpha \rangle =|\alpha e^{i\psi } \rangle $ and 
\begin{eqnarray}
  U_\theta |\alpha \rangle |\delta \rangle =
|\alpha \cos \theta + \delta \sin \theta \rangle | \delta \cos\theta
-\alpha \sin\theta \rangle 
  \;,
\end{eqnarray}
we have 
\begin{eqnarray}
&&\chi (\lambda ,\mu ) = \int \frac {d^2 \alpha } {\pi N_A} \int \frac {d^2 \gamma } {\pi
    N_B}   e^{-\frac {|\alpha 
      |^2}{N_A}-\frac {|\gamma
      |^2}{N_B}} \label{9}
  \\& & \times  
  \langle \alpha \cos \theta + \gamma \sin \theta |
  \alpha \cos \theta + \gamma e^{i(\lambda -\mu)\omega _A -i \lambda
    \omega _B}\sin \theta \rangle \nonumber \\& & \times  
  \langle \gamma  \cos \theta - \alpha \sin \theta |
  \gamma  \cos \theta - \alpha e^{i \lambda \omega _B -i (\lambda -\mu
    )\omega _A}\sin \theta \rangle
\;.\nonumber 
\end{eqnarray}
Finally, from the relation
\begin{eqnarray}
\langle \alpha | \gamma \rangle =\exp \left (-\frac 12
|\alpha |^2 -\frac 12 |\gamma |^2 +\bar \alpha \gamma \right )
\;
\end{eqnarray}
and lengthy but straightforward Gaussian integration we obtain
\begin{eqnarray}
&&  \chi (\lambda ,\mu )
  =\big\{1- \sin ^2 \theta
 \times \label{chilmu}\\& & 
  \left [ (N_A+N_B +2 N_A N_B)[\cos (\mu \omega _A -
    \lambda (\omega _A -\omega _B)) -1] \right.
\nonumber \\& &
\left.  +i (N_A -N_B)\sin (\mu \omega _A -
\lambda (\omega _A -\omega _B))\right ] \big \}^{-1} \;.\nonumber 
\end{eqnarray}
We easily check the identity $\chi [i\beta _B , i(\beta _B -\beta _A)
]=1$, corresponding to the standard fluctuation theorem. Indeed, the time-reversal
symmetry of the unitary operation provides the stronger identity 
$  \chi [i\beta _B - \lambda , i(\beta _B -\beta _A) -\mu  ]=\chi (
\lambda ,  \mu )$, 
corresponding to the Gallavotti-Cohen microreversibility
\cite{evans,gal}, and equivalent to the detailed fluctuation theorem
\cite{andrie,cth,sini,frq}
\begin{eqnarray}
\frac{p(W,Q_H)}{p(-W,-Q_H)}= e^{(\beta _B- \beta _A )Q_H +\beta _B W}\;.\label{dett}
\end{eqnarray}
Notice the symmetry
$  \langle W^n Q_H ^m \rangle
=
     \left ( \frac {\omega _A} {\omega _B -\omega
       _A}\right ) ^m \langle W^{n+m} \rangle $ 
and, from the first law,
$\langle Q_C ^n \rangle = (-\omega
_B / \omega _A)^n \langle Q_H ^n \rangle $. 
\par Using Eqs. (\ref{mom}) and (\ref{chilmu}) one obtains the
following averages and variances of work and heat
\begin{eqnarray}
  &&\langle W \rangle =  (\omega _A -\omega _B)(N_B -N_A)\sin ^2
  \theta\;,\\
  &&\langle Q_H \rangle = \omega _A (N_A -N_B)\sin ^2
  \theta = \frac {\omega _A}{\omega _B -\omega _A} \langle W \rangle
  \;,\\
  &&\mbox{var}(W)
  = (\omega _A -\omega _B)^2[N_A +N_B +2
    N_A N_B  \nonumber \\& &
    \qquad \ \ \ + (N_A-N_B)^2 \sin ^2 \theta ]\sin ^2
  \theta\;,\\ &&\mbox{var}(Q_H) = 
  \frac {\omega ^2_A}{(\omega _A -\omega _B)^2} \mbox{var}(W)
  \;,\\ && \mbox{cov}(W,Q_H) 
  =\frac {\omega _A}{\omega _B -\omega _A} \mbox{var}(W) \;.  
\end{eqnarray}
We can identify three regimes of operation, namely
\begin{eqnarray}
  a)&&\ \ \ \omega _A >  \omega _B  \quad \& \quad
  N_A > N_B \qquad \quad \mbox{\ \ \ \ heat engine},
  \nonumber \\ b)&&\ \ \  
\omega _A >  \omega _B \quad  \& \quad N_A <  N_B \qquad \quad 
\mbox{\ \ \ \ refrigerator},
  \nonumber \\c)&&\ \ \ 
\omega _A <  \omega _B  \quad (\implies  N_A  >  N_B ) \ \ \ \ 
\mbox{thermal accelerator},
\nonumber
\end{eqnarray}
where correspondingly we have
\begin{eqnarray}
a)&&\ \ \ \ \langle W \rangle <0,\qquad  \langle Q_H 
\rangle >0,\qquad \langle Q_C \rangle <0 \,;\nonumber \\
b)&&\ \ \ \ \langle W \rangle > 0,\qquad  \langle Q_H \rangle <0,
\qquad \langle Q_C \rangle > 0 \,; \nonumber \\
c)&&\ \ \ \ \langle W \rangle >0,\qquad  \langle Q_H \rangle >0,
\qquad \langle Q_C \rangle <0\,.
\nonumber
\end{eqnarray}
We notice that for both the heat engine and the refrigerator the sign of
$\mbox{cov}(W,Q_H)$ is negative. On the other hand, for the thermal
accelerator where external work is consumed to increase the heat flow
from hot to cold reservoir the covariance is positive. 
In terms of the temperature of the reservoirs, it is useful to observe
that
\begin{eqnarray}
  \beta _A \omega _A \leq \beta _B \omega _B  \iff N_A \geq N_B
  \,,
\label{ifff}
\end{eqnarray}
and thus the three regimes are equivalently identified by
\begin{eqnarray}
a)\ \ \frac{T_B}{T_A}<\frac{\omega _B}{\omega _A}<1; 
\ \ \ \ b)\ \ \frac{\omega _B}{\omega _A}< \frac{T_B}{T_A}
<1;\ \ \ \ c)\ \frac{\omega _B}{\omega _A}>1.
\;\nonumber
\end{eqnarray}
\begin{figure}[htb]
{\includegraphics[scale=0.55]{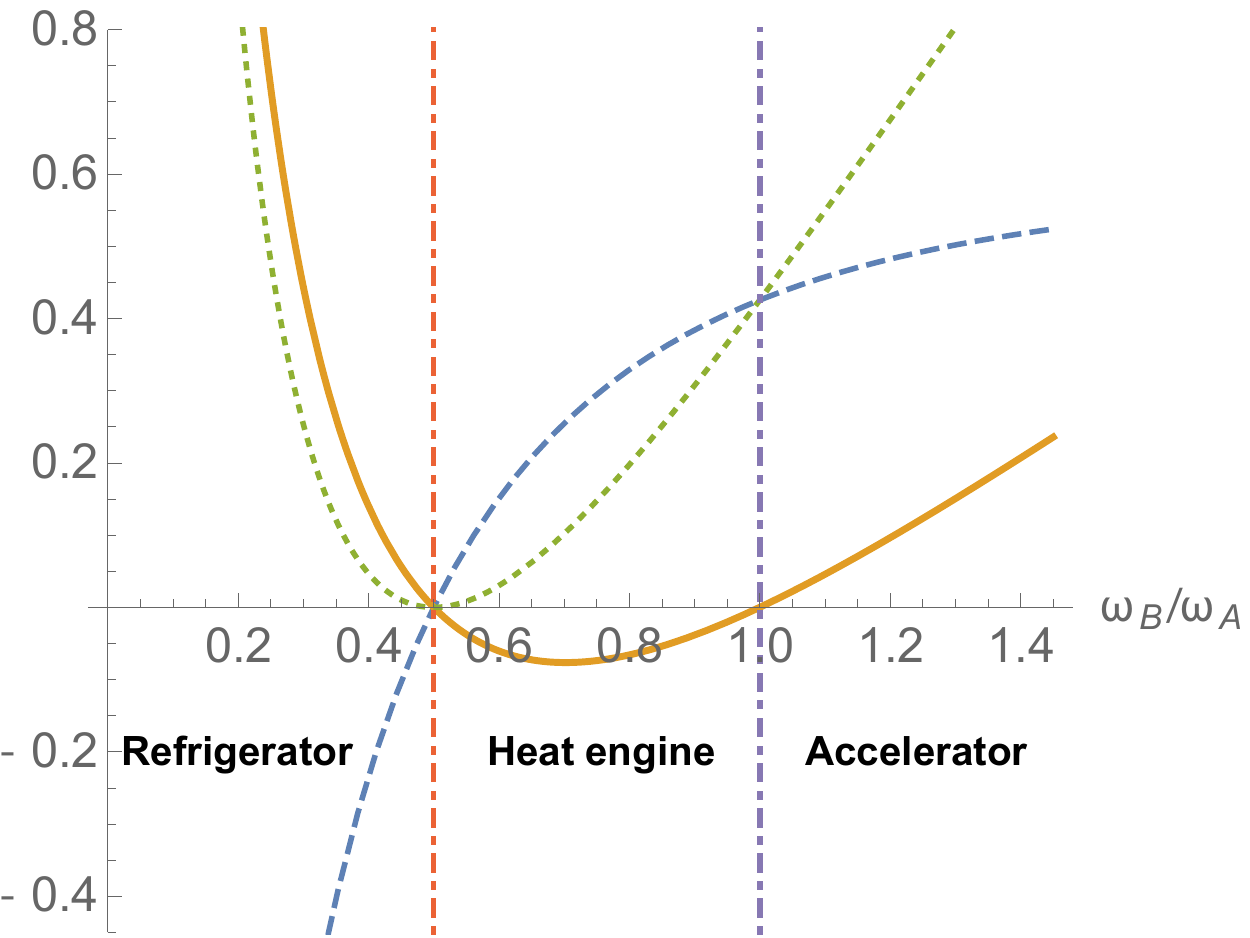}}
\caption{Plot of work, heat and entropy production
  (thick, dashed, and dotted style, respectively) for $\omega _A =1$, $\beta _A =1$, $\beta
  _B=2$, and $\theta =\pi/2$ versus the
  ratio $\omega _B/ \omega _A$, in their three regions of operation.}
\end{figure}
\par The efficiency of the heat engine is given by
\begin{eqnarray}
\eta  = \frac{\langle -
  W \rangle }{\langle Q_H \rangle } =1- \frac{\omega _B}{\omega _A}
\leq 1 - \frac {T_B}{T_A}\equiv \eta _C\,,
\end{eqnarray}
corresponding to the Otto
cycle efficiency. The Carnot efficiency $\eta _C$ is achieved only for
$\omega _A/\omega _B =T_A/T_B$ (i.e., for $N_A=N_B$ with zero output
work).  Analogously, the coefficient of performance (COP) for the
refrigerator is given by
\begin{eqnarray}
\zeta =\frac{\langle Q_C \rangle }{\langle W
  \rangle }=\frac{\omega _B}{\omega _A -\omega _B}\leq \frac {T_B}{T_A
  -T_B}=\zeta _C\,.
\;
\end{eqnarray}
Notice that both the efficiency and the COP are independent
of $\theta $ and the temperature of the reservoirs.
\par Since $[U_\theta
  , a^\dag a + b^\dag b]=0$ one has $\Delta E_b = -\frac {\omega
  _B}{\omega _A}\Delta E_a$, and hence the entropy production $\langle
\Sigma \rangle $ can be written as follows
\begin{eqnarray}
   \langle \Sigma \rangle &&
  =\beta _A \Delta E_a +\beta _B \Delta E_b =
\frac{\beta _A \omega _A -\beta _B \omega _B}
  {\omega _A- \omega
    _B}
  \langle W \rangle 
  \nonumber
  \\& &=
(\beta _A \omega _A -\beta _B \omega _B)  (N_B -N_A)\sin ^2 \theta 
  \;.\label{epro}
\end{eqnarray}
From Eq. (\ref{ifff}), as expected, one always has $\langle \Sigma
\rangle \geq 0$. Work, heat and entropy production are depicted in Fig. 2 for 
parameters $\omega _A=1$, $\beta _A =1$, and $\beta _B=2$, with
$\theta =\pi/2$. 
\par By  the identity
$\frac{\beta _A \omega _A -\beta _B \omega _B}
  {\omega _A- \omega
    _B}= -\frac{1}{T_B}(\frac{\eta _C}{\eta }-1)$,
  for the heat engine one obtains the relation 
\begin{eqnarray}
   \langle \Sigma \rangle = \frac {\langle -W \rangle }{T_B} \left (\frac
           {\eta _C}{\eta }- 1 \right )\;\label{epp}
\end{eqnarray}
between average extracted work, entropy
production and efficiency. 
Analogously, for the refrigerator 
one has
\begin{eqnarray}
\langle \Sigma \rangle = \frac {\langle Q_C \rangle }{T_A} \left (\frac{1}{\zeta}
-\frac{1}{\zeta _C}\right )\;.
\end{eqnarray}
\section{Thermodynamic uncertainty relations}
Using Eqs. (11-15) one can obtain 
the inverse signal-to-noise ratios
\begin{eqnarray}
\frac{\mbox{var}(W)}{\langle W \rangle
  ^2}
&&
=
\frac{\mbox{var}(Q_H)}{\langle Q_H \rangle ^2}=\frac
     {\mbox{cov}(W,Q_H)}
     {\langle W \rangle \langle Q_H \rangle }
    \nonumber \\& &
     =
\frac{N_A +N_B +2 N_A N_B}{(N_A -N_B)^2 \sin ^2 \theta}+1
\;.\label{invw}
\end{eqnarray}
These ratios are minimized versus $\theta $
for $\theta =\frac \pi 2$, for which also the entropy production
$\langle \Sigma \rangle $ achieves the maximum.  Notice also that
operating at zero entropy production (i.e. for $N_A \rightarrow N_B$,
thus approaching the Carnot efficiency) will produce a divergence in
Eq. (\ref{invw}).  By combining
Eqs. (\ref{epro}) and (\ref{invw}), independently of $\theta $ we
obtain the following exact relation 
\begin{eqnarray}
\frac{\mbox{var}(W)}{\langle W \rangle ^2} =  \frac {h(\beta _A \omega _A
  -\beta _B \omega _B)}{\langle \Sigma \rangle}+1  
\;,\label{reduc}
\end{eqnarray}
where $h(x)=x \,\mbox{cth} (x/2)$. Then, reducing the noise-to-signal
ratio associated to work extraction (or cooling performance) comes at
a price of increased entropy production.  Since $h(x)\geq 2$, the
following thermodynamic uncertainty relation is always satisfied
\begin{eqnarray} 
\frac{\mbox{var}(W)}{\langle W \rangle ^2} \geq \frac {2}{\langle \Sigma \rangle}+1  
\;,\label{bb}
\end{eqnarray}
and then also the standard TUR $\mbox{var}(W)/ \langle W \rangle
    ^2 \geq 2 /\langle \Sigma \rangle $. 
\par\noindent In Fig. 3 we plot the work variance and compares it with the bound
obtained by Eq. (\ref{bb}), for fixed parameters $\omega _A=1$, $\beta
_A =1$, and $\beta _B=2$. Differently from the two-qubit case studied
in Ref. \cite{timpa}, we do not observe a violation of the standard
TUR. Indeed, the tightest saturable bound from Ref. \cite{timpa} 
\begin{eqnarray}
\frac{\mbox{var}(W)}{\langle W \rangle ^2} \geq f(\langle \Sigma
\rangle )
  \;,\label{loo}
\end{eqnarray}
where $f(x)=\mbox{csch}^2[g(x/2)]$ and $g(x)$ denotes the inverse
function of $x\, \mbox{tanh}(x)$, becomes quite loose for the present
bosonic engine for $\omega _B \ll \omega _A$ (see Fig. 3). For a more
direct comparison with the two-qubit engine, where the standard TUR
can be violated, see Appendix A. The effect of finite thermalization
times on the TUR is also considered in Appendix D.
\par From Eqs. (\ref{epp}) and (\ref{bb}) we can write a relation
  between the average extracted work, fluctuations and efficiency 
\begin{eqnarray}
\langle -W \rangle \leq \frac {\mbox{var}(W)}{2 T_B} \left (\frac {\eta
  _C} {\eta} -1 \right )  
\;.\label{eqpo}
\end{eqnarray}
\begin{figure}[!tb]
{\includegraphics[scale=0.51]{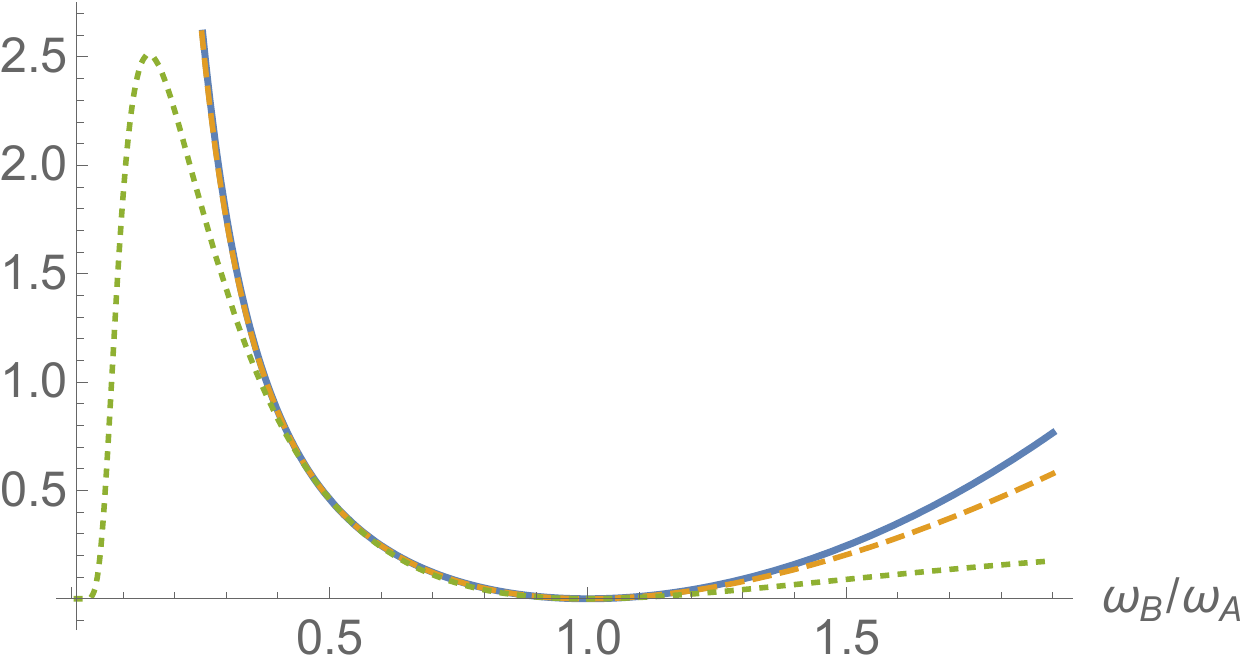}}
\caption{Plot of the work variance $\mbox{var}(W)$
  (thick style) and the function
  $\langle W \rangle ^2 \left ( \frac {2}{\langle \Sigma
    \rangle}+1\right )$ in
  dashed style, for $\omega _A =1$, $\beta _A =1$, and $\beta _B=2$ versus the
  ratio $\omega _B/ \omega _A$. 
The dotted curve is obtained by the lower bound in Eq. (\ref{loo}) derived
in Ref. \cite{timpa}.}
\end{figure}
\par\noindent This can also be written as a bound on the efficiency,
determined by the average work and fluctuations, 
namely
\begin{eqnarray}
\eta \leq \frac{\eta _C}{1+ 2T_B \langle -W \rangle / \mbox{var}(W)}\;. \label{eqpo2}
\end{eqnarray}
We notice that Eqs. (\ref{eqpo}) and (\ref{eqpo2}) are analogous
to the universal trade-off derived in
Ref. \cite{piet} for steady-state engines permanently coupled to heat
baths.  The bound (\ref{eqpo2}) shows that in order to increase
the efficiency, one must either sacrifice the output work or increase
the fluctuations, thus decreasing the engine reliability.
\par We observe that both the stochastic work and heat come as integer multiple
of $\omega _A -\omega _B$ and $\omega _A$, respectively. In fact, this
can also be understood \cite{luk,sele} by noting that the characteristic
function has periodicity $\frac{2 \pi}{|\omega _A -\omega _B|}$ and
$\frac{2\pi }{\omega _A}$ in the variables $\lambda $ and $\mu $.
The joint probability for work and heat is then given by 
\begin{eqnarray}
  &&\!\!
  p[W=m(\omega _A -\omega _B), Q_H = n \omega _A]= \frac {\omega _A |\omega
    _A -\omega _B|}{(2 \pi)^2} \label{jwh} \\& &
\!\!\!\!  \times \int _{-\frac{\pi}{|\omega _A -\omega _B |}}^
{\frac{\pi}{|\omega _A -\omega _B|}} d \lambda
\int _{-\frac{\pi}{\omega _A }}^
     {\frac{\pi}{\omega _A}} d \mu \, \chi (\lambda ,\mu ) 
     e^{-i \lambda m (\omega _A -\omega _B) -i \mu n \omega _A}
\nonumber \\& &   \!\!\!\!   = 
     p[W=m(\omega _A -\omega _B)] \delta _{n,-m}  = 
p[Q_H=n \omega _A]\delta _{m,-n} \nonumber
  \;,
\end{eqnarray}
where, by the derivation given in Appendix B, 
\begin{widetext}
\begin{eqnarray}
& &
  p[Q_H =n\omega _A]=p[W=-n(\omega _A -\omega _B)]=
  \frac{1}{\sqrt{1+ 2
      (N_A+N_B+2N_A N_B) \sin^2 \theta+
(N_A-N_B)^2\sin ^4\theta }} \label{closed}
\\& & \times
\left\{
\begin{array}{ll}
  \left (\frac{1+ 
    (N_A+N_B+2 N_A N_B)\sin ^2 \theta
  - \sqrt{1+ 2(N_A+N_B+2 N_A N_B)\sin ^2 \theta +(N_A-N_B)^2\sin ^4
    \theta}}{2 N_B (N_A+1)\sin ^2 \theta}\right )^{n} & \qquad \mbox{for }n\geq 0 
  \;,\\
\left (\frac{1+ (N_A+N_B+2 N_A N_B)\sin ^2 \theta
  - \sqrt{1+ 2(N_A+N_B+2 N_A N_B)\sin ^2 \theta +(N_A-N_B)^2\sin ^4
    \theta}}{2 N_A (N_B+1) \sin ^2 \theta }\right )^{|n|} & \qquad
\mbox{for }n < 0\;. 
\end{array}
\right.
\nonumber
\;
\end{eqnarray}
\begin{figure}[htb]
\begin{subfigure}
  \centering
\includegraphics[scale=0.5]{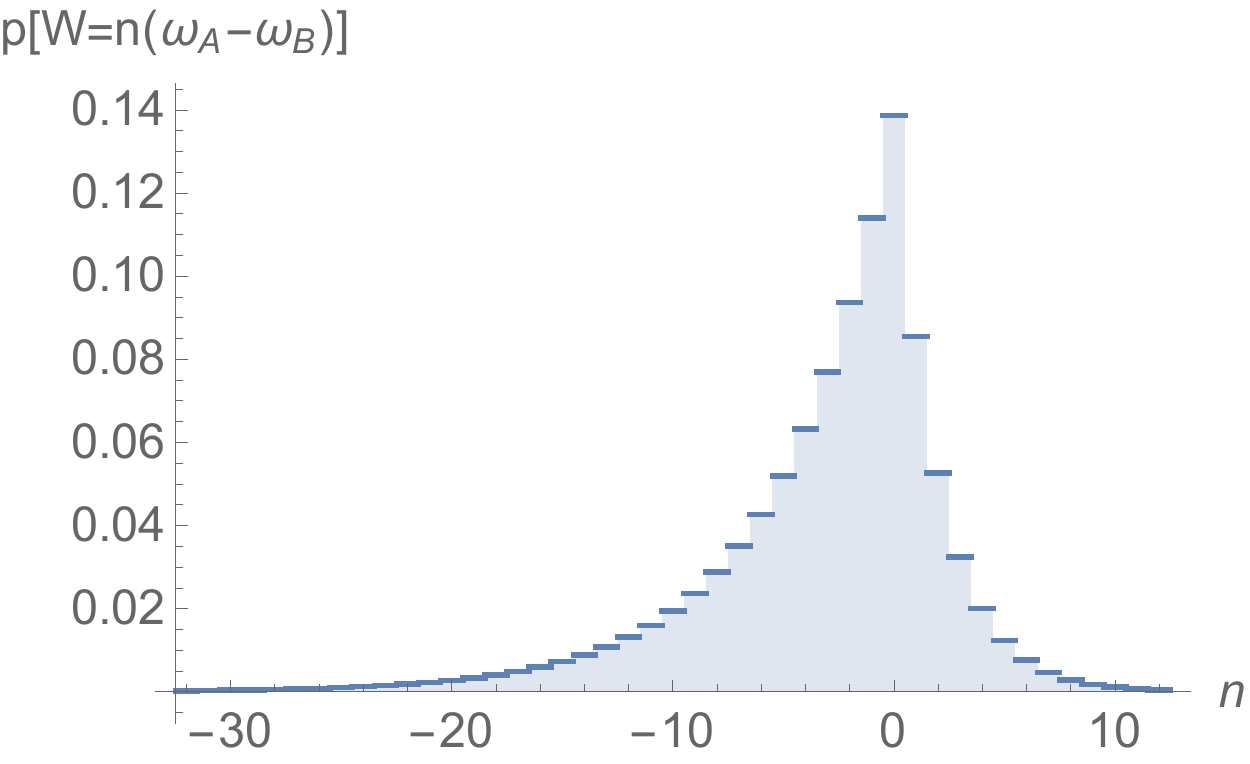}
\end{subfigure}
\ \ \ \ \ \ \ \ \ \ \ \ \ \ \ 
\begin{subfigure}%{0.5\textwidth}
  \centering
  % include second image
\includegraphics[scale=0.5]{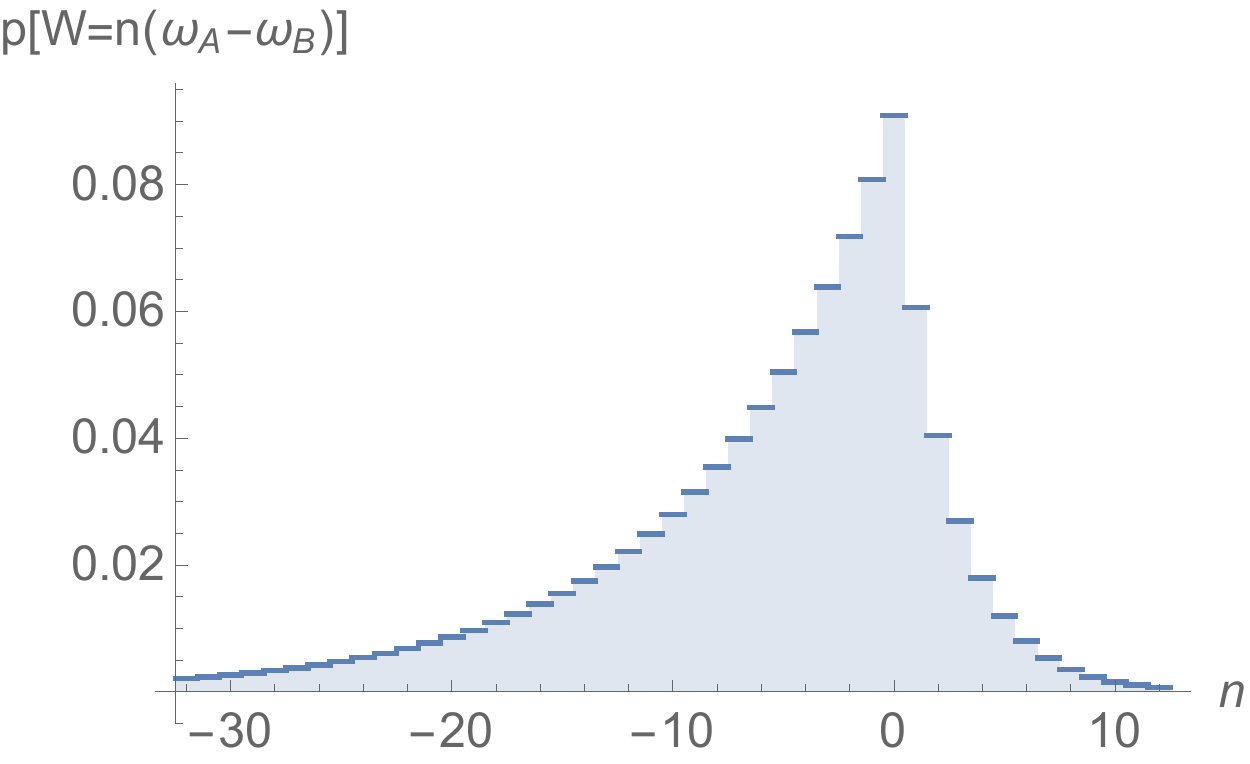}  
\end{subfigure} 
\caption{Distribution of the extracted work in $\omega _A -\omega _B$
  units, for $N_A=8$ and $N_B=2$, for interaction strength $\theta
  =\pi/4$ (left) and $\theta =\pi/2$
  (right). By exchanging $n \rightarrow -n$, the same histograms
  represent the probability of heat released by the hotter reservoir in $\omega _A $
  units [see Eq. (31) and (B7)].}
\end{figure}
\end{widetext}
\par In Fig. 4 we report the work probability for $N_A=8$ and $N_B=2$,
pertaining to two different values of strength interaction, i.e. $\theta
=\pi/4$ and $\theta =\pi/2$.
\par From the form of Eq. (\ref{jwh}), similarly to the case of the
two-qubit swap engine \cite{cpf}, one recognizes that the efficiency
is indeed a self-averaging quantity. In fact, in principle the
efficiency $\eta = \frac {\langle -W \rangle }{\langle Q_H \rangle }$
is different from the expectation of the stochastic efficiency $\eta
_s = \langle -W /Q_H \rangle $.  However, here we have for all moments
\begin{eqnarray}
\langle (-W/Q_H)^n \rangle =\langle -W/Q_H \rangle ^n =\left ( 1-
\frac {\omega _B }{\omega _A}\right )^n\;,
\end{eqnarray}
namely there are no
efficiency fluctuations.
\par The closed form for the probability of Eq. (\ref{closed}) allows one
to explicitly verify the detailed fluctuation theorem in  Eq. (\ref{dett})
as follows
\begin{eqnarray}
&&  \frac{p[W=-n(\omega _A -\omega _B),Q_H =n\omega _A]}
    {p[W=n(\omega _A -\omega _B),Q_H =-n\omega _A]}=
%  \frac{p[Q_H =n\omega _A]}
%       {p[Q_H =-n\omega _A]}
       \left [\frac{N_A (N_B+1)}{N_B (N_A
           +1)}\right ]^n
      \nonumber \\& &
%       =e^{(\beta_B\omega _B -\beta _A \omega _A)n}
    =
  e^{(\beta _B -\beta _A)n \omega _A - \beta _B n (\omega _A -\omega
    _B)}=
e^{(\beta _B- \beta _A )Q_H +\beta _B W}\;.
\end{eqnarray}
In Appendix C we provide a general discussion on the special character
of the joint probability $p(W,Q_H)$ and the outline of the
generalization of the present approach to study Otto engines with
alternative unitary interactions. 
\section{Conclusions}
In conclusion, by adopting the two-point-measurement protocol for the
joint estimation of work and heat, we have derived exact expressions
for work and heat fluctuations pertaining to two-mode bosonic Otto
engines, where two quantum harmonic oscillators are alternately
subject to a tunable unitary bilinear interaction and to thermal
relaxation to their own reservoirs. We have derived the characteristic
function for work and heat, and obtained the full joint probability 
of the stochastic work and heat.
\par The presented thermodynamic
uncertainty relations show the interdependence among average extracted
work, fluctuations and entropy production, which hold in all range of
coupling parameter between the two quantum harmonic oscillators.  Our
results confirm the general meaning of TURs, namely that reducing the
noise-to-signal ratio associated with a given current comes at a price
of increased entropy production.
\par The direct derivation of TURs by
explicit measurement protocols can be effective in a variety of stroke
thermodynamic engines. Within this approach, the relevance of the
algebraic properties of the interactions naturally emerges.
\par The connection between fluctuation theorems, estimation protocols and
thermodynamic uncertainty relations represents a significant advance
in our understanding of nonequilibrium phenomena, and is relevant for
the design of quantum thermodynamic machines, by posing strict bounds
that relate work, heat, fluctuations, efficiency, and reliability.

%\begin{widetext}
\onecolumngrid
\appendix
\section{a comparison with the two-qubit Otto engine}
It is interesting to compare the results for the two-mode bosonic Otto
engine with the case of the two-qubit Otto engine. Hence, we extend the study of
Ref. \cite{timpa} to the case of partial swap, by considering a two-qubit unitary
interaction 
\begin{eqnarray}
  U_\theta = \left (
  \begin{array}{cccc}
    1 & 0 & 0 & 0 \\
    0 & \cos \theta & \sin \theta &0
    \\
    0 & -\sin \theta & \cos \theta &0 \\
    0 & 0 & 0 & 1
    \end{array}
    \right )\;,
\end{eqnarray}
  where we used the tensor-product ordered basis
  $|0 0 \rangle ,|0 1 \rangle ,|1 0 \rangle ,|1 1 \rangle $ for two
  qubits. The characteristic function is still obtained by
  Eq. (\ref{6})  of
  the main text, where now $H_X = - \omega _X |0 \rangle \langle
  0|$. A simple calculation gives 
\begin{eqnarray}
  &&  \chi(\lambda ,\mu )=1 + \nonumber \\& &
  \sin ^2 \theta \{(N_A+N_B-2 N_A N_B)[\cos
    (\mu
    \omega _A -\lambda (\omega _A -\omega _B))-1]
    +i (N_A-N_B)[\sin    (\mu
    \omega _A -\lambda (\omega _A -\omega _B))] \}
  \;,
\end{eqnarray}
where now $N_X=(e^{\beta _X \omega _X}+1)^{-1}$. The odd and even moments are given by
\begin{eqnarray}
  &&\langle Q_H   ^{2n+1} \rangle =\omega _A ^{2n+1}
  (N_A
-N_B) \sin^2 \theta \;,\\& &
\langle Q_H ^{2n} \rangle =\omega _A ^{2n}(N_A
+N_B -2 N_A N_B) \sin^2 \theta
\;,
\end{eqnarray}
and  $  \langle W^n Q_H ^m \rangle  =  \left ( \frac {\omega _B -\omega _A} {\omega
  _A}\right ) ^n \langle Q_H ^{n+m} \rangle $.  
The entropy production has the same formal expression of the bosonic
case, namely
\begin{eqnarray}
  \langle \Sigma \rangle
  = (\beta _A \omega _A -\beta _B \omega _B)
  (N_B -N_A)\sin ^2 \theta 
  \;,\label{epro2}
\end{eqnarray}
whereas the inverse signal-to-noise ratios reads 
\begin{eqnarray}
\frac{\mbox{var}(W)}{\langle W \rangle
  ^2}
= \frac{\mbox{var}(Q_H)}{\langle Q_H \rangle ^2}=
\frac{N_A +N_B -2 N_A N_B}{(N_A -N_B)^2 \sin ^2 \theta}-1
\;.\label{invw2}
\end{eqnarray}
For the qubit engine, Eq. (\ref{reduc}) of the main text is then replaced
with
\begin{eqnarray}
  \frac{\mbox{var}(W)}{\langle W \rangle ^2}
  = \frac {h(\beta _A \omega _A
  -\beta _B \omega _B)}{\langle \Sigma \rangle}-1  
\;,
\end{eqnarray}
where, remarkably, the same function $h(x)=x \,\mbox{cth} (x/2)$
appears. Since around the affinity $x=\beta _A \omega
_A -\beta _B \omega _B$ one has $2 \leq h(x) \leq 2 + \frac{x^2}{6}$
the standard TUR
\begin{eqnarray} 
\frac{\mbox{var}(W)}{\langle W \rangle ^2} \geq \frac {2}{\langle \Sigma \rangle}  
\;\label{bb2}
\end{eqnarray}    
can be tinily violated for the qubit engine, as shown in
Ref. \cite{timpa}.  
In Fig. 5 we report the signal-to-noise ratio $ \langle W \rangle ^2
/\mbox{var}(W)$ along with the function $\langle \Sigma \rangle /2$
for the cases $\theta =\pi/2$ and $\theta =\pi /3$.  We observe that
the region of violation of the thermodynamic uncertainty relation
(\ref{bb2}) is shrunk for decreasing values of $\theta$.
\begin{figure}[ht]
\begin{subfigure}%{0.5\textwidth}
  \centering
  % include first image
  \includegraphics[scale=0.5]{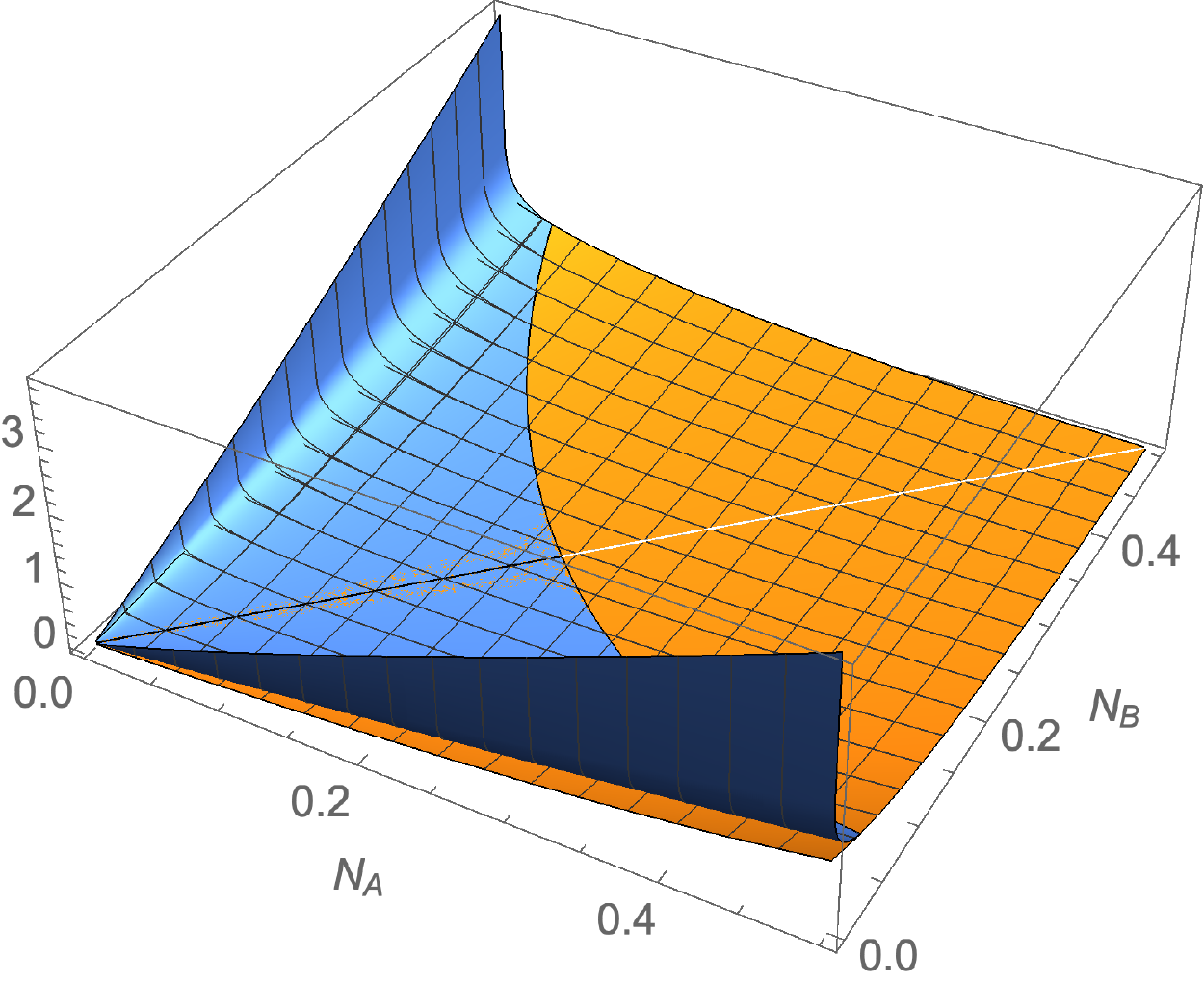}
\end{subfigure}
\ \ \ \ \ \ \ \ \ \ \ \ \ \ \ 
\begin{subfigure}%{0.5\textwidth}
  \centering
  % include second image
  \includegraphics[scale=0.5]{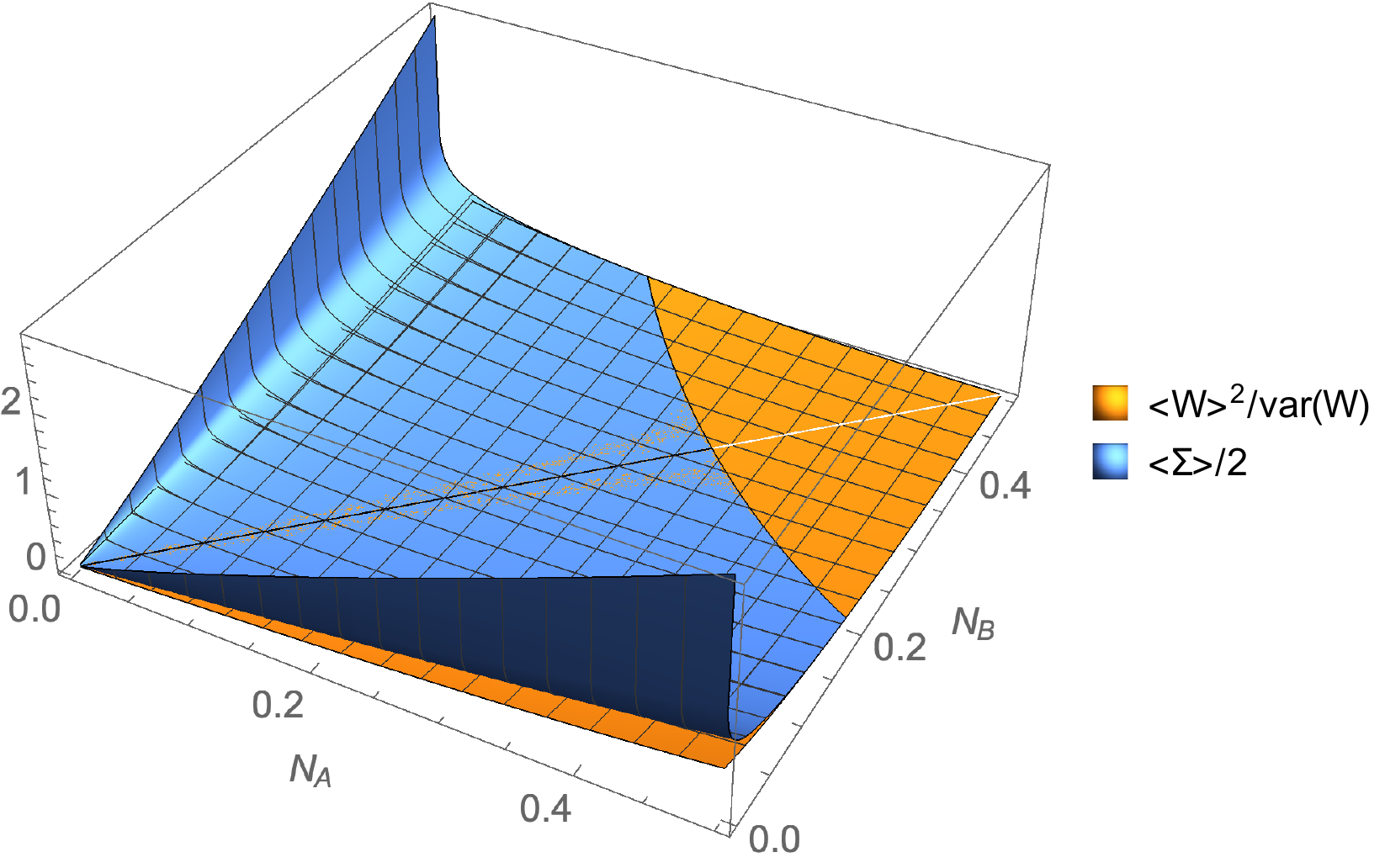}  
\end{subfigure}
\caption{Plot of the signal-to-noise ratio of work
  $  \langle W
  \rangle ^2 /\mbox{var}(W)$ and scaled entropy production $\langle \Sigma
    \rangle /2$ for the qubit Otto engine with $\theta =\pi/2$ (left) and
    $\theta =\pi /3$ (right) as a function of parameters $N_A$ and
    $N_B$.} 
\end{figure} 
\par For the qubit engine the probability for the
stochastic heat and work has finite outcomes and is obtained as follows
\begin{eqnarray}
&&  p[Q_H= n \omega _A]=p[W=-n(\omega _A -\omega _B) ] \nonumber \\& &
  =\frac {1}{2\pi}
  \int _{0} ^ {2 \pi}
\left \{1 + 
\sin ^2 \theta \left [(N_A+N_B-2 N_A N_B)[(\cos
    \mu )-1]
    +i (N_A-N_B)\sin  \mu \right ]
 \right  \}
  \, e^{-i\mu n}\, d\mu
\nonumber \\& & =
\left\{
\begin{array}{ll}
  1- (N_A +N_B -2 N_A N_B) \sin^2 \theta & \qquad \mbox{for }n=0 
\;,\\ 
N_A (1-N_B) \sin^2 \theta & \qquad \mbox{for }n=1\;,\\
N_B (1-N_A) \sin^2 \theta
& \qquad \mbox{for }n=-1\;.\\  
\end{array}
\right.
\;
\end{eqnarray}
As we have shown above, this three-point probability  may
give rise to a violation of Eq. (\ref{bb2}).  The finiteness of the
stochastic outcomes and the different algebra of operators concur to
provide a different thermodynamic uncertainty relation with respect to
the bosonic case.  We recall that the saturable bound of the main text
(\ref{loo})  provides a stronger violation of the standard TUR and is achieved
by a two-point distribution, as shown in Ref. \cite{timpa}.
\section{probability for the stochastic work and heat of the bosonic Otto engine}
From the Eq. (\ref{jwh}) of the main text,
in order to obtain the probability for the stochastic work and heat we
need to perform the following integral
\begin{eqnarray}  p[Q_H =n\omega _A]&&=p[W=-n(\omega _A -\omega _B)]  
\\&&  = \frac{1}{2\pi} \int _{0}^{2\pi}
\big\{1- 
\left [ (N_A+N_B +2 N_A N_B)
           [(\cos 
    \mu  ) -1]    +i (N_A -N_B)\sin \mu  \right ] \sin ^2 \theta  \big \}^{-1} e^{-i
      \mu  n} d \mu \;.\nonumber 
\end{eqnarray}
The
integral can be solved by using the residue theorem, after posing $z=e^{i
  \mu}$ and integrating on the complex plane
along the unit circle $\gamma $, with $d \mu  = dz /(iz)$. Then, we
have
\begin{eqnarray}
p[Q_H =n\omega _A]&&   
  = \frac{1}{2\pi} \int _\gamma 
  \big\{1- 
  \left [ (N_A+N_B +2 N_A N_B) 
[(z + z^{-1})/2 -1]    +i (N_A -N_B) (z-
z^{-1})/(2i) \right ] \sin ^2 \theta \big \}^{-1} z^{-n} \frac {dz} {iz}\nonumber \\& &
=\frac{1}{2\pi i} \int _\gamma 
\frac{z^{-n}}
     {[1+
         (N_A+N_B +2N_A N_B)\sin^2 \theta] z -
            [N_A(N_B +1)z^2 +N_B(N_A+1)]       \sin ^2\theta}dz 
\;.
\end{eqnarray}
For $n\leq 0$ the poles are easily evaluated as
\begin{eqnarray}
  z_\pm = \frac{1+ 
    (N_A+N_B+2 N_A N_B) \sin ^2 \theta
    \pm \sqrt{1+ 2
      (N_A+N_B+2 N_A N_B)\sin ^2 \theta +(N_A-N_B)^2\sin ^4
      \theta}}{2 
    N_A (N_B+1) \sin ^2 \theta}
\;.
\end{eqnarray}
We observe that
\begin{eqnarray}
  z_+ > \frac{1+ 
    [(N_A+N_B+2 N_A N_B) + |N_A -N_B|]\sin ^2 \theta}
  {2     N_A (N_B+1)\sin ^2 \theta} \;.
\end{eqnarray}
Then, for $N_A \geq N_B$ clearly one has $z_+>1$. For $N_A < N_B$, one
also has
\begin{eqnarray}
  z_+ > \frac{1+ 2N_B (N_A +1) \sin ^2 \theta}
  {2     N_A (N_B+1) \sin ^2 \theta} >1\;,
  \;
\end{eqnarray}
since  $N_B > N_A >0  \iff N_B
(N_A +1)> N_A (N_B+1)$.  Hence, the pole $z_+$ lies outside the
unitary circle.
\par The residue for the first-order pole $z_-$ is given
by
\begin{eqnarray}
&&  \mbox{Res} \left ( \frac {z^{|n|}}
  {[1+
      (N_A+N_B +2N_A N_B)\sin^2 \theta] z - 
    [N_A(N_B
    +1)z^2 +N_B(N_A+1)] \sin ^2\theta }, z_- \right ) \nonumber \\& &
= \left. \frac{z^{|n|}}
{[1+
    (N_A+N_B +2N_A N_B)\sin^2 \theta]  - 2 
  N_A(N_B
    +1)z \sin ^2\theta}\right | _{z=z_-}
\nonumber \\& &  = \frac{1}{\sqrt{1+ 2
    (N_A+N_B+2N_A N_B)\sin^2 \theta +
(N_A-N_B)^2\sin ^4\theta }}
\nonumber \\& & \times \left ( \frac{1+ 
  (N_A+N_B+2 N_A N_B)\sin ^2 \theta
  - \sqrt{1+ 2
    (N_A+N_B+2 N_A N_B) \sin ^2 \theta +(N_A-N_B)^2\sin ^4
    \theta}}{2 
  N_A (N_B+1) \sin ^2 \theta}\right )^{|n|}
\;
\end{eqnarray}
For $n >0$, we also have a $n$-order pole in $z=0$. However, we can recast
the integration as for the case $n<0$ by the change of variable
$\mu \rightarrow - \mu $, which is then equivalent to exchange
$N_A$ with $N_B$. Hence, one obtains the closed expression for
the probability for the stochastic work and
heat of Eq. (\ref{closed}). 
\par In the
case of the swap engine $\theta = \frac \pi 2$, one can directly
derive the analytic expression for $p[Q_H=n\omega _A ]$ as follows
\begin{eqnarray}
  &&
  p[Q_H=n \omega_A ]=
p[W=-n(\omega _A -\omega _B)]=
  \sum_{l,s =0}^\infty
  \Tr [(|l \rangle \langle l|
  \otimes I_B) U_{\pi /2}(|s \rangle \langle s| \otimes \rho _{N_B} )
  U^\dag _{\pi/ 2}]   \langle s|\rho _{N_A} |s \rangle \,
  \delta_{n,s-l}
\nonumber \\& & 
=\sum _{l,s=0}^\infty \frac{1}{N_A+1}\left( \frac{N_A}{N_A+1} \right
)^s\frac{1}{N_B+1}\left( \frac{N_B}{N_B+1} \right
)^l   \delta_{n,s-l}
=
\left\{
\begin{array}{ll}
  \frac{1}{1+N_A+N_B} \left (\frac{N_A}{N_A
  +1}\right )^n  & \qquad \mbox{for }n\geq 0 
\;,\\ 
 \frac{1}{1+N_A+N_B} \left (\frac{N_B}{N_B
  +1}\right )^{|n|}   & \qquad \mbox{for }n< 0\;, 
\end{array}
\right.
\end{eqnarray}
consistent with Eq. (\ref{closed}) for $\theta = \frac \pi 2$. 
\section{general consideration on the joint probability $p(W,Q_H)$.}
We would like to make some general considerations about the special
character of the joint probability $p(W,Q_H)$.  Let us come back to
the characteristic function $\chi (\lambda ,\mu )$ in Eq. (\ref{6}) of the
main text. We
notice that the periodicity in $\lambda $ and $\mu $ which is evident
in Eq. (\ref{chilmu}) can be indeed recognized from the expression of
Eq. (\ref{6}) 
without explicit calculation, but exploiting the algebra of bosonic
operators, since one can rewrite
\begin{eqnarray}
  \chi (\lambda ,\mu )
  = \Tr [U^\dag _\theta U_{\xi } \rho _0]
\;,\label{alge}
\end{eqnarray}
where $\xi= \theta e^{i
  \lambda (\omega _A -\omega _B)- i\mu \omega _A}$.
The fact that $\chi (\lambda ,\mu )$ is
a function of the single variable $\lambda (\omega _A -\omega _B)-\mu
\omega _A$ is due to the symmetry $[U_\theta ,a^\dag a + b^\dag b]=0$,
and from this the Kronecker delta is obtained as
\begin{eqnarray}
  p[W=m(\omega _A -\omega _B), Q_H = n \omega _A]&&= \frac {\omega _A |\omega
    _A -\omega _B|}{(2 \pi)^2} 
\int _{-\frac{\pi}{|\omega _A -\omega _B |}}^
{\frac{\pi}{|\omega _A -\omega _B|}} d \lambda
\int _{-\frac{\pi}{\omega _A }}^
     {\frac{\pi}{\omega _A}} d \mu \, \chi (\lambda ,\mu ) 
     e^{-i \lambda m (\omega _A -\omega _B) -i \mu n \omega _A}
     \nonumber \\& &    
  = \delta _{m,-n} \, \frac{1}{2\pi} \int _{0}^{2\pi} \chi \left (0, \frac{\mu}{\omega
    _A} \right ) e^{-i
             \mu  n} d \mu
     \;.
\end{eqnarray}
This feature can also be obtained in other thermodynamic engines where 
a different observable is a constant of motion during the unitary
strokes. For example,  one can consider the unitary $V_\theta =\exp
(\theta a^\dag b^2 - \theta ^* a b^{\dag 2})$, where now the constant of
motion is $2 a^\dag a + b^\dag b$. The characteristic
function is then given by $\chi(\lambda ,\mu )=\Tr [V^\dag _\theta
  V_\zeta \rho_0]$ with $\zeta = \theta e^{i
  \lambda (\omega _A -2\omega _B)- i\mu \omega _A}$, and hence
\begin{eqnarray}
  &&  p[W=m(\omega _A -2 \omega _B), Q_H = n \omega _A]=p[W=m(\omega
    _A -2 \omega _B)] \delta _{n,-m}= 
  p[Q_H= n \omega _A]  \delta _{m,-n}\;.
\end{eqnarray}
Clearly, also in this case the efficiency $\eta = \langle -W /Q_H \rangle =
1-2 \omega _B/\omega _A$ has no fluctuations. Even without finding 
explicitly the stochastic distribution one can exploit this result for
proving some thermodynamic properties.
For example, in this case we can write the average entropy production as follows
\begin{eqnarray}
  \langle \Sigma \rangle = - 
   \frac{\beta _A \omega
    _A -2 \omega _B \beta _B}{\omega _A} \langle Q_H \rangle 
  =    \frac{\beta _A \omega
    _A -2 \omega _B \beta _B}{\omega _A -2 \omega _B} \langle W
  \rangle \;. 
\end{eqnarray}
By requiring the positivity of the entropy production one can easily
infer the condition for having a heat-engine operation $\langle Q_H
\rangle > 0$ and $\langle W \rangle <0$, namely $\beta _A \omega _A <
2 \beta _B \omega _B$ and $\omega _A > 2 \omega _B$. We notice that
the first of these conditions is equivalent to $N_A > N_B^2/(2 N_B
+1)$.  
Further work is required in order to obtain other properties related to
higher moments (e.g. thermodynamic uncertainty relations), since the
algebra of operators $(a^\dag  b^{2}, a  b^{\dag 2}, a^\dag a, b^\dag b)$ is
not closed. The presented approach might be fruitful for the study of
nonlinear optical interactions from a thermodynamic perspective.
\par Similarly, for the two-mode squeezing unitary interaction
$S_r =\exp [r (a^\dag b^\dag -  a b)]$ for
which $[S_r,a^\dag a - b^\dag b]=0$, one obtains
\begin{eqnarray}
  &&  p[W=m(\omega _A + \omega _B), Q_H = n \omega _A]=
p[W=m(\omega
  _A + \omega _B)] \delta _{n,-m}=    
p[Q_H= n \omega _A] \delta _{m,-n} \;.
\end{eqnarray}
In this case the engine can work just as a dud machine, since one
always has $\langle W \rangle \geq 0$, along with $\langle Q_H \rangle
, \langle Q_C \rangle \leq 0$. Basically, in this case the unitary
strokes perform work $W=(\omega _A+ \omega _B)
(N_A+N_B+1)\sinh^2 r$ to build correlations that are then converted
to heat when the two harmonic
oscillators relax to equilibrium by their thermal reservoirs. 
This is consistent with a general result
obtained in Ref. \cite{bil}, where it is shown that the presence of
initial correlations is needed to extract work by the interaction
$S_r$. By exploiting the closed algebraic transformations
\begin{eqnarray}
  &&S^\dag _r a S_r
  =a \cosh r +
  b^\dag  \sinh r ,\label{sq mode a-1}\\
&& S^\dag _r b S_r
  =b \cosh r   + 
  a^\dag  \sinh r,\label{sq mode b-1}
\end{eqnarray}
from the general formula of the main text (\ref{mom}) one obtains
\begin{eqnarray}
&&\langle W \rangle = (\omega _A +\omega _B)(N_A +N_B +1)\sinh ^2
  r\;,\nonumber \\& &
\frac{\mbox{var}(W)}{\langle W \rangle ^2}= \frac{N_A+N_B+2N_A
  N_B+1}{(N_A+N_B+1)^2 \sinh^2 r} +1\,.  
\end{eqnarray}
The entropy production reads
\begin{eqnarray}
  \langle \Sigma \rangle = \frac
          {\beta _A \omega _A +\beta _B \omega _B}
          {\omega _A +\omega _B} \langle W \rangle
  \;,
\end{eqnarray}
and hence, for any value of the interaction strength $r$, one obtains the exact relation 
\begin{eqnarray}
  \frac{\mbox{var}(W)}{\langle W \rangle ^2}=
  \frac{h(\beta _A \omega _A + \beta _B \omega _B)}{\langle \Sigma \rangle } +1
\;.
\end{eqnarray}
Remarkably, as for the interaction $U_\theta$, the function
$h(x)=x \,\mbox{cth}(x/2)$ appears, and then also in this case the thermodynamic
uncertainty relation $\mbox{var}(W) /\langle W \rangle ^2 \geq 2/\langle
\Sigma \rangle +  1$ holds.

By an analogous derivation of Eq. (\ref{closed}) given in Appendix B, one can obtain
the probability for the stochastic work and
heat as 
\begin{eqnarray}
& &
  p[Q_H =n\omega _A]=p[W=-n(\omega _A +\omega _B)]=
  \frac{1}{\sqrt{1+ 2
      (N_A+N_B+2N_A N_B+1) \sinh^2 r +
(N_A+N_B+1)^2\sinh ^4 r  }} \nonumber 
\\& & \times
\left\{
\begin{array}{ll}
\left (\frac{1+ \sinh ^2 r (N_A+N_B+2 N_A N_B+1)
  - \sqrt{1+ 2 (N_A+N_B+2 N_A N_B+1)\sinh^2 r +(N_A+N_B+1)^2\sinh ^4
    r}}{2 (N_A+1)(N_B+1) \sinh^2 r }\right )^{n} & \qquad \mbox{for }n\geq 0 
  \;,\\
\left (\frac{1+ (N_A+N_B+2 N_A N_B+1)\sinh^2 r
  - \sqrt{1+ 2 (N_A+N_B+2 N_A N_B+1) \sinh^2 r +(N_A+N_B+1)^2\sinh ^4
    r}}{2 N_A N_B \sinh^2 r}\right )^{|n|} & \qquad
\mbox{for }n < 0\;. 
\end{array}
\right.
\label{closed2}
\;
\end{eqnarray}
\par A further interesting observation comes from the specific form of
the stochastic distributions of Eqs. (\ref{closed}) and
(\ref{closed2}), namely an asymmetric Bose-Einstein distribution over
$n \in \mathbb Z$. This is due to the property of the interactions
$U_\theta $ and $S_r$ of transforming initial Gibbs states in a final
correlated state which locally (i.e. the two partial traces on each
mode after the interaction) is still of the Gibbs form. In fact, from
the perspective of pure probability theory such power-law expressions
along with the detailed fluctuation theorem generally give rise to the
thermodynamic uncertainty relation $\mbox{var}(W)/\langle W \rangle
^2=\mbox{var}(Q_H)/\langle Q_H\rangle ^2 \geq 2/\langle \Sigma 
\rangle +1 $, as shown in the following.  Let us assume a general
stochastic distribution over $n \in \mathbb Z$ of the form
\begin{eqnarray}
p[Q_H=n v]= p[W=n k]=\left\{
\begin{array}{ll}
  \alpha  x^n
& \qquad \mbox{for }n\geq 0 \,,
  \;\\
  \alpha  y^{|n|} 
      & \qquad
\mbox{for }n < 0 \,, 
\end{array}
\right.
\end{eqnarray}
with arbitrary real $v$ and $k$, and with $x$ and $y \in [0,1]$.  
The normalization condition of probability implies $\alpha
=(1-x)(1-y)/(1- xy)$. One easily obtains the identities 
\begin{eqnarray}
&&\langle W \rangle = \frac {k}{v} \langle Q_H \rangle = \frac
        {k}{(v+k)\beta _B -h \beta _A}\langle \Sigma
        \rangle =k\frac {x-y}{(1-x)(1-y)}\;, \\& & 
 \frac{\mbox{var}(W)}{\langle W \rangle ^2
        }=\frac{\mbox{var}(Q_H)}{\langle Q_H \rangle }=
         \frac{(x+y)(1-x)(1-y)}{(x-y)^2} +1= \frac{(x+y) [(v+k)\beta
            _B -v \beta _A]}{(x-y)\langle \Sigma \rangle} +1\;.\label{turgen} 
\end{eqnarray}
The detailed fluctuation theorem $ \frac{p[W=n k]}{p[W=- nk]}=e^\Sigma $
also provides the constraint $x/y=e^{(v+k)\beta _B -v \beta _A}$. Then
Eq. (\ref{turgen}) rewrites as 
\begin{eqnarray}
\frac{\mbox{var}(W)}{\langle W \rangle ^2
        }=\frac{\mbox{var}(Q_H)}{\langle Q_H \rangle ^2}=
 \frac{h[(v+k)\beta _B -v \beta _A]}{\langle \Sigma \rangle} +1 \geq  \frac
        {2}{\langle \Sigma \rangle  } +1 \;.\label{turgen2} 
\end{eqnarray}
\section{partial thermalization for the bosonic swap engine}
The study of the case of partial thermalization requires some care,
for two different reasons. First, one has to ignore a transient time
in order to consider the possible stabilization of a periodic steady state at
the beginning of each cycle. Second, for general coupling parameter 
$\theta $ the resulting state at the beginning of each cycle, even in
the periodic steady-state regime, is a correlated state which does not
commute with $H_A$ and $H_B$, and hence the approach of the two-point
measurement scheme to obtain the characteristic function is not
justified. This second issue, however, does not affect the engine in
the case of perfect swap $\theta =\frac \pi 2$, since in any case the
initial state at each cycle is of bi-Gibbsian form, and we can study
partial thermalization as follows. 
Let us consider the usual bosonic dissipation described by a Lindblad
master equation to model thermalization \cite{carm}, namely
\begin{eqnarray}
  \dot{\rho }= \gamma _A (N_A+1) \left (a \rho a^\dag  -\frac 12 a^\dag a \rho
    - \frac 12 \rho a^\dag a \right )+
    \gamma _A N_A \left (a^\dag \rho a -\frac 12  a a^\dag \rho
    -\frac 12  \rho a  a^\dag \right )
    \;,\label{lind}
\end{eqnarray}
and analogously for mode $b$. For simplicity let us assume 
equal damping rates $\gamma _A=\gamma _B \equiv \gamma $ for both
modes.  At the end of the $(n+1)$-th cycle with finite thermalization
time $\tau $ the state will be bi-Gibbsian with mean occupation
numbers  
\begin{eqnarray}
  &&N_A ^{n+1}= e^{-\gamma \tau } N_B^{n} + (1-e^{-\gamma
    \tau})N_A\;,\label{big}  \\
&& N_B ^{n+1}= e^{-\gamma \tau }N_A^{n}  + (1-e^{-\gamma \tau})N_B\;.\label{big2}  
\end{eqnarray}
After transient time, the cycles lead to a periodic state
corresponding to the steady-solution of Eqs. (\ref{big}) and (\ref{big2}), which are given by
\begin{eqnarray}
&&\tilde N_A= (N_A + e^{-\gamma \tau }N_B)/(1+e^{-\gamma \tau
  })\;,\nonumber \\& &
\tilde N_B= (N_B + e^{-\gamma \tau }N_A)/(1+e^{-\gamma \tau })\;.  
\end{eqnarray}
It follows that the characteristic function is still given by
Eq. (\ref{chilmu}) of the main
text, along with the
replacement of $N_A$ and $N_B$ with $\tilde N_A$ and $\tilde N_B$,
respectively. Then, the average work, heat and entropy production per
cycle give in Eqs. (13), (14) and (21), respectively,
are just rescaled by the factor $\mbox{tanh}(\gamma \tau /2)$. The
effect of partial thermalization is more involved for physical
quantities related to higher moments. For example, Eq. (\ref{invw}) for the inverse
signal-to-noise ratios is replaced with
\begin{eqnarray}
  \frac {\mbox{var}(W)}{\langle W \rangle ^2}=\frac{(1+e^
    {-2 \gamma \tau})[N_A(N_A+1)+N_B (N_B+1)]+ 2 e^{-\gamma
      \tau }(N_A+N_B+2 N_A N_B)}{(1-e^{- \gamma \tau })^2 (N_A -N_B)^2}
\;.\label{turqu}
\end{eqnarray}
Clearly, for $\tau \rightarrow +\infty$, Eq. (\ref{invw}) is recovered. In
Fig. 6 we plot the signal-to-noise ratio for fixed value of the
parameter $N_A=3$ versus varying $N_B$, for different values of
$\gamma \tau$, where it is apparent the detrimental effect of
decreasing the thermalization times.
\begin{figure}[ht]
  \includegraphics[scale=0.58]{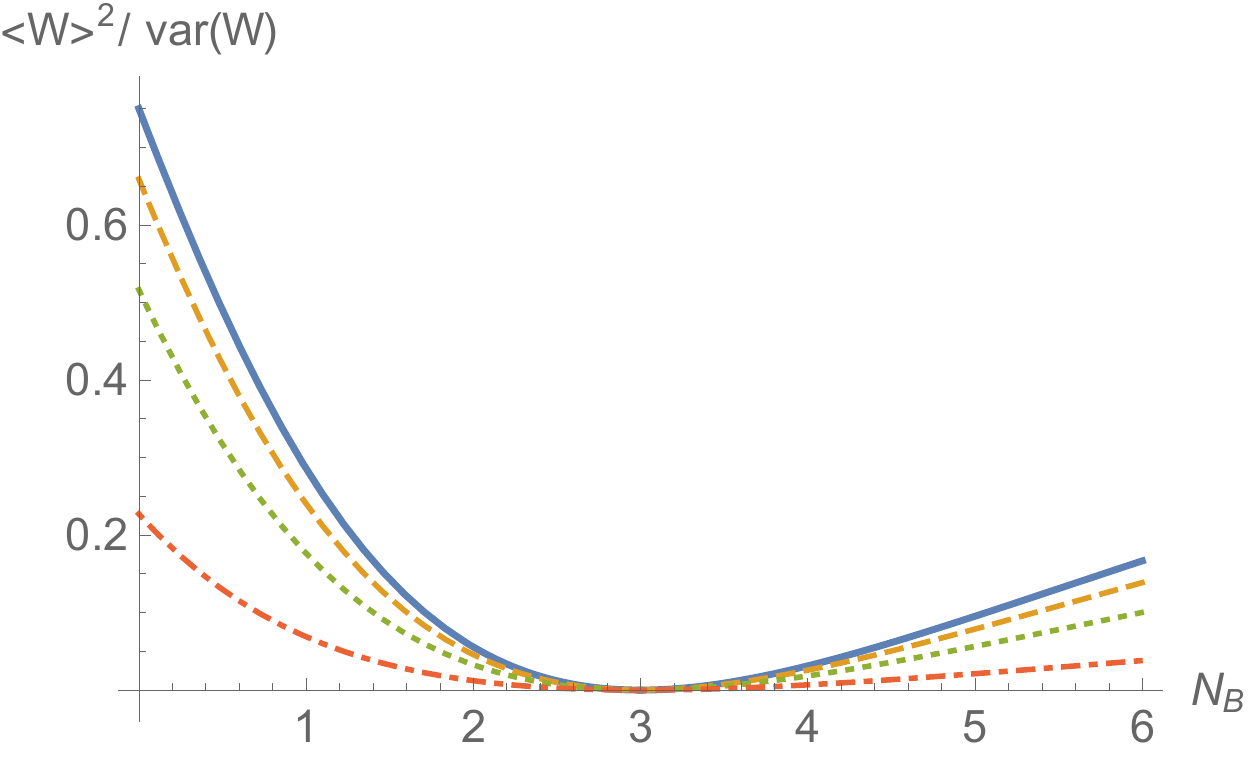}
\caption{Signal-to-noise ratio of the work for the bosonic swap engine ($\theta =\frac \pi
  2$) with $N_A=3$ versus occupation number $N_B$ for ideal
  thermalization (solid), and finite thermalization times $\gamma \tau
  =3.,2.$ and $1.$ (dashed, dotted and  dot-dashed, respectively).}
\end{figure}
\par The function $h(\beta _A \omega _A - \beta _B \omega _B)$ in
Eq. (\ref{reduc}) of the main text is replaced with
\begin{eqnarray}
&&  v(\beta _A \omega _A,\beta _B \omega _B,\gamma \tau)
  \equiv  (\beta _A \omega _A - \beta _B \omega _B)
  \\& & \times   \frac{
    (1+e^{-\gamma
      \tau})^2\cosh (\beta _A \omega _A)+(1+e^{-\gamma
      \tau})^2\cosh (\beta _B \omega _B) -(1+e^{-2\gamma \tau})
    \cosh (\beta _A \omega _A - \beta _B \omega _B) -e^{-\gamma \tau
    }(4+e^{-\gamma \tau})-1
  }{(1-e^{-2\gamma \tau})[\sinh (\beta _A \omega _A)-\sinh (\beta _B
      \omega _B)-\sinh   (\beta _A \omega _A - \beta _B \omega _B)]
  }    \nonumber
\;.
\end{eqnarray}
One can easily prove the bound
\begin{eqnarray}
  v(\beta _A \omega _A,\beta _B \omega _B,\gamma \tau)\geq 2 \coth
  (\gamma \tau /2)
\;,
\end{eqnarray}
and hence the thermodynamic uncertainty relation
\begin{eqnarray} 
\frac{\mbox{var}(W)}{\langle W \rangle ^2} \geq  \frac {2}{\langle
  \Sigma \rangle}\coth (\gamma \tau /2) +1
\;.\label{bbtau}
\end{eqnarray}
This bound shows that thermodynamic uncertainty relations can be
informative also for more realistic engines where finite
thermalization times are considered. Partial thermalization clearly
affects the signal-to-noise ratio of the extracted work. When treating
specific microscopic interactions via time-dependent Hamiltonian or
assigning a time cost to the unitary transformations, one
may study optimal time allocation between thermalization strokes and
unitary strokes in order to maximize the extracted work at non-zero
power.
\par The replacement rule $(N_A,N_B)\rightarrow (\tilde N_A,
\tilde N_B)$ also applies to the joint probability of the
stochastic work and heat. This implies that even in the case of partial
thermalization the efficiency for the swap engine remains a
non-fluctuating quantity.  We notice, however, that a detailed
fluctuation theorem as in Eq. (\ref{dett}) holds provided that $\beta _A$ and
$\beta _B$ are replaced by the effective inverse temperatures $\tilde
\beta _X =\frac {1}{\omega _X}\ln \left ( \frac{\tilde N_X +1}{\tilde
  N_X}\right )$.

\par For arbitrary interaction parameter $\theta $,  we argue that the
issue of the presence of correlations or coherence in the periodic steady
states could be addressed by replacing the two-measurement protocol
with a full-counting-statistics approach, along the lines of
Ref. \cite{soli}.

\twocolumngrid

%\end{widetext}

\end{document}